\documentstyle{l-aa}
\input psfig.sty

\begin{document}

   \thesaurus{08         % A&A Section 6: Form. struct. and evolut. of stars
              (08.01.1;  % Cosmogony,
               08.16.4;  % Planets and satellites: general,
               08.03.4;  % Solar system: general,
               08.12.1;)}  % Stars: formation of,

\title{Carbon molecules in the inner wind of the oxygen-rich Mira IK Tauri}

%\subtitle{I. Overviewing the $\kappa$-mechanism}

   \author{Debiprosad Duari\inst{1} \and Isabelle Cherchneff\inst{1} \and Karen Willacy\inst{2}
%          \inst{1}
          }

   \offprints{dpd@saturn.phy.umist.ac.uk}

   \institute{Department of Physics, UMIST, PO Box 88, 
Manchester, M60 1QD, UK \and 
	 Jet Propulsion Laboratory, Pasadena CA 91109, USA}

   \date{Received 1998; accepted  1998}

   \maketitle
\markboth{D. Duari et al.: Carbon molecules in Ik Tau}{}

\begin{abstract}
The gas-phase non-equilibrium chemistry of the inner wind of the oxygen-rich 
Mira variable IK Tau (NML Tau) is investigated using a physio-chemical 
model describing the periodically shocked gas close to the 
stellar photosphere. We predict  the formation in large amounts 
of a few carbon species in the inner envelope, in particular 
CO$_2$, which has been recently detected with ISO in the spectra 
of several oxygen-rich semi-regular and Mira stars. The theoretical abundances 
are also in excellent agreement with values derived from 
millimeter and sub-millimeter observations, pointing to the fact that some 
carbon species in oxygen stars do form from shock chemistry in the
inner layers and travel the envelope as ``parent'' species.

   \end{abstract}

%________________________________________________________________

\section{Introduction}

Carbon-bearing molecules have been identified in the outer envelopes 
of many oxygen-rich (O-rich), late-type stars at millimeter and 
microwave wavelengths (Deguchi \& Goldsmith 1985, Lindqvist et al 
1988, Omont et al. 1993, Bujarrabal et al. 1994) and include HCN, 
 CS, OCS, HNC and CN. Because HNC is typically  
formed in the outer envelopes of AGB stars  from ion-molecule 
chemistry triggered by the penetration of cosmic rays and ultra-violet 
radiation, it was first thought that the observed carbon species could  
be produced in the outer wind of oxygen stars. Photo-chemical models 
of several oxygen Miras by Willacy \& Millar (1997, hereafter WM97) 
tried to reproduce the observations and relied on the injection of 
methane, CH$_4$, ammonia, NH$_3$, silicon sulfide, SiS and hydrogen 
sulfide, H$_2$S, to generate a carbon-rich chemistry at large radii. 
These input molecular abundances are questionable since there exist 
no observational or theoretical evidences for the formation of these 
species in the inner and intermediate envelopes of O-rich Miras. The 
model succeeded in reproducing the observed values of certain species  
but failed to reproduce some molecular abundances, in particular that 
of hydrogen cyanide, HCN, which is known to be  a ``parent'' molecule 
in carbon stars and forms in the inner regions of the winds (Willacy 
\& Cherchneff 1998, hereafter WC98).  

The recent ISO detection of several infrared (IR) emission bands of 
CO$_2$ in various O-rich Miras and semi-regular stars by Justtanont 
et al. (1998) suggests that this carbon species has to form in the 
deep layers of the stellar winds in order to excite the IR transitions 
observed.  In this letter, we investigate the non-equilibrium chemistry 
of the inner regions close to the stellar photosphere of the oxygen Mira 
IK Tau (NML Tau) and raise the following questions: (a) Can some of 
these carbon molecules form in the
inner parts of the wind and be ejected as ``parent'' species ? ; (b) 
Can methane and ammonia form in the inner wind
of O-rich Mira stars ?

%__________________________________________________________________

\section{A model for the inner wind:} 

IK Tau is a variable O-rich Mira of spectral type M6. The 
stellar parameters considered in this study are listed in Table 
1.  Its distance was derived by Le Sidaner \& Le Bertre (1994) 
to be 220 pc while its pulsation period is $\sim$ 470 days (Hale 
et al. 1997). Le Sidaner \& Le Bertre assumed a stellar temperature 
of 2000 K for their radiative transfer model of the IR excess, in 
good agreement with the value used in this study. In order to derive 
a radius for the star we assumed a luminosity close to the  canonical 
value for typical Miras stars given by the Period-Luminosity 
relation derived by Feast (1996). The derived radius satisfies the 
standard pulsation equation for Miras of Fox \& Wood (1982)  
%--------------------
\begin{equation}
Q=P(M/M_{\odot})^{0.5}\,(R/R_{\odot})^{-1.5}
\end{equation}
%---------------------
where we assume that the star is pulsating in its fundamental 
mode (Q$=0.09$). This gives a stellar mass of 1 M$_\odot$, 
typical of O-rich Miras variables. There exists observational 
evidence for O-rich Miras to pulsate in their first overtone 
(Feast 1996) and we will study the impact of this result on our 
stellar and chemical model in a forthcoming paper.   

%%__________________________________________________ One column table
\begin{table}
\begin{flushleft}
\caption{\label{tab:params} IK Tau - stellar parameters:
$\gamma$ and $\alpha$ are  defined as in Cherchneff et al. (1992) and  
X (Y) $\equiv {\rm X \times 10^Y}$. }
\begin{tabular}{llll}
\hline
T$_{\star}$  & 2100 K  & L$_\star$ & 2.0 $\times$ 10$^3$ L$_\odot$\\
R$_\star$    & 305 R$_\odot$  & M$_\star$ & 1 M$_\odot$\\
P            & 470 days  & n(r$_{\rm shock}$)  & 3.7 (15) cm$^{-3}$\\
$r_{\rm shock}$        & 1.0 R$_\star$   & $\alpha$  & 0.6\\
$\gamma =v_{\rm shock}/v_{\rm esc}$ & 0.89 & C/O ratio & 0.75 \\
\hline
\end{tabular}
\end{flushleft}
\end{table}
%-----------------------------------------------------

We consider the inner wind as a narrow region above the photosphere
which experiences the passage of strong, periodic shocks generated
by stellar pulsation. The model describes both the immediate
region (thermal cooling region) and the hydrodynamical cooling region
of the post-shock structure as described by Fox \& Wood (1985),
Bertschinger \& Chevalier (1985) and WC98. 
More details on the model for the inner winds of AGB stars are given in 
Cherchneff et al. (1998). In our model of IK Tau, we choose a 
shock speed v$_{\rm shock} = 32$ km s$^{-1}$ in agreement with shock velocities
derived from the CO IR line analysis of Hinkle et al. (1997). 
The periodic shocks steepen at r$_{\rm shock}= 1$ 
R$_\star$ in our model and levitate the nearby gas layers to larger
radii, these regions falling down to their initial position because of
stellar gravity. This periodic motion generates a pattern of gas
excursions as illustrated in Figure 1.  
The physical parameters characterising the shocked, inner wind 
are listed in Table 2. 

%%----------------------------------------------------------- S_vib
\begin{figure}
\caption{Gas excursions induced by an initial shock of 32 km s$^{-1}$. The arrows
show the shock position.} 
%[htbp]
     \psfig{figure=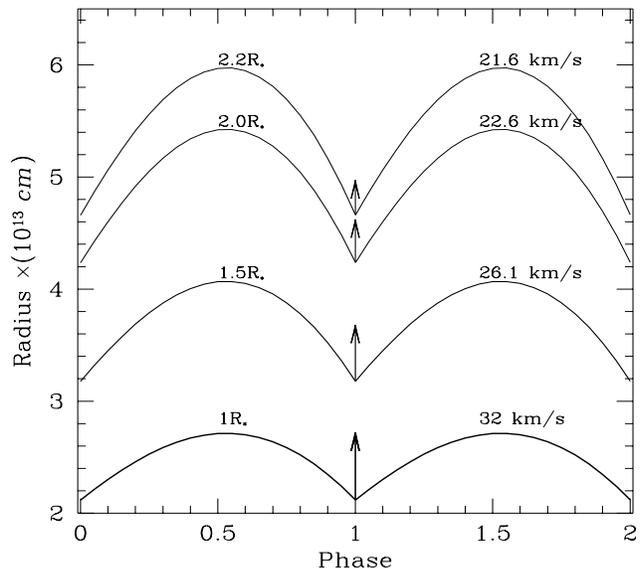,height=8cm,width=8.8cm,angle=0}
%      \caption{Vibrational stability equation of state
%               $S_{\rm vib}(\lg e, \lg \rho)$.
%               $>0$ means vibrational stability
%              }
%         \label{FigVibStab}
\end{figure}
%%
%%
%%  14.Sep.'90: Demo Vs.
%%______________________________________________________________
 
%------------------------------------------------------------------------
\begin{table*}
\begin{flushleft}
\caption{\label{tab:jump}Pre-shock, shock front and excursion ($\equiv$ post-shock) gas temperature and number density
as a function of position in the envelope and shock strengths. M is the Mach number associated with each shock speed.} 

\begin{tabular}{ccccccccc}
\hline
Position & Shock Vel. & M &\multicolumn{2}{c}{Pre-shock} & \multicolumn{2}{c}{
Shock Front} & \multicolumn{2}{c}{Start of excursion}  \\
 ($R_\star$) & (km s$^{-1}$) &{\rm } &$T_0$ (K) & $n_0$ (cm$^{-3}$) & $T$ (K) & $n$ (cm$^{-3}$)
& $T$ (K) & $n$ (cm$^{-3}$) \\
\hline
 1.0 &32.0 &  10.6 & 2100.0 &3.62 (15) & 47861 & 2.08 (16) & 6271 & 1.32 (17) \\ 
1.5 & 26.1  &  9.74 &1646.5& 1.49 (13) & 31924 & 8.49 (13) & 4711 & 4.80 (14) \\
2.0 & 22.6 & 9.19  & 1385.5 & 5.02 (11) & 24059  & 2.84 (12) & 3838 & 1.49 (13) \\
2.2 & 21.6 &  9.02  &1308.5 & 1.78 (11) & 21934 & 1.01 (12) & 3572 & 5.13 (12) \\
\hline
\end{tabular}
\end{flushleft}
\end{table*}
 
\section{The chemistry:}

      We consider 68 chemical species and 752 reactions and include all
possible chemical routes at work in a dense gas, that is,  termolecular 
and bimolecular neutral-neutral reactions. No photo-processes (dissociation 
or ionisation) are considered as the UV stellar radiation field is low 
for the effective temperature quoted in Table 1. Also, we do not consider 
radiative processes in the post-shock gas on the ground of the theoretical 
models of post-shock structures by Fox \& Wood (1985). Indeed, for a 
molecular, cool pre-shock gas, they find that the ionization level in 
the thermal cooling region is low and we consider the collisional 
dissociation of molecular hydrogen to be the dominant coolant in the 
immediate post-shock region (WC 1998, Cherchneff et al. 1998). 

The reaction rates are taken from the RATE95 UMIST database (Millar 
et al. 1997), Baulch et al. (1992), Cherchneff et al. (1992), Mick 
et al. (1994) and the NIST database (Mallard et al. 1994). Details about 
assumptions involved in the chemistry used in the model are given by WC98.

We assume thermal equilibrium (TE) for the stellar photosphere and derive 
molecular abundances for the effective temperature, gas number density 
and C/O ratio given in Table 1 for IK Tau. We then ``shock'' the 
photosphere and investigate the chemistry in the immediate cooling 
layer and the hydrodynamical cooling part ($\equiv$ excursion) of the 
post-shock region. The model (the immediate post-shock region followed 
by excursion at one position in the envelope) is run over two pulsation 
cycles to check for periodicity. The output abundances for one model 
are then used as the input to that for the shock at the next distance and 
are rescaled according to the local gas number density.

%-------------------------------------------------------------------------
   \begin{table*}
      \caption{Calculated fractional abundances (relative to the total 
gas number density) versus shock strength and radius. Data from 
millimeter observations are listed with the following references: 
1 - Bujarrabal et al. (1994), 2 - Lindqvist et al. (1988), 
3 - Omont et al. (1993), 4 - Justtanont et al. (1998) :
{\it in other O-rich miras. No observations for IK Tau} , 
5 - Menten \& Alcolea (1995).}
%%%      \[ 
%%%         \begin{array}{p{0.5\linewidth}llllll}
%%%	\hline
\begin{flushleft} 
\begin{tabular}{llllllcc} 
\hline\noalign{\smallskip}
 Species  &   T.E.  & 32 km s$^{-1}$ & 26.1  km s$^{-1}$ & 22.6  km s$^{-1}$ & 21.6  km s$^{-1}$ & \multicolumn{2}{c}{Observational values} \\  
  & 1. R$_{\star}$ & 1. R$_{\star}$ &  1.5 R$_{\star}$ & 2. R$_{\star}$ &  2.2 R$_{\star}$ &  &  \\
            \noalign{\smallskip}
            \hline
            \noalign{\smallskip}
  H   &     6.65 (-2)    &   1.08 (-3)    &   6.82 (-3)    &   
1.85 (-1)    &   3.19 (-1)   & \cr
H$_2$  &     6.52 (-1)    &   7.08 (-1)    &   7.03 (-1)     
&   5.51 (-1)    &   4.36 (-1)   & \\
  Si  &     1.03 (-8)    &   7.85 (-13)    &   4.42 (-16)    
&   4.19 (-22)    &   2.63 (-22)   & \\
  S   &     1.37 (-5)    &   1.20 (-6)    &   7.92 (-6)    
&   1.96 (-5)    &   1.84 (-5)   & \\
  C$_2$H$_2$ &     8.56 (-17)    &   5.99 (-12)    &   6.05 (-15)    
&   4.59 (-17)    &   6.52 (-12)   & \\
 {\bf  CS }   &  {\bf  1.32 (-10)}   & {\bf   8.09 (-6) }  & {\bf   1.14 (-5) }   
& {\bf  5.15 (-7) }   & {\bf   2.75 (-7) }  &  1.0 (-7)$^1$ & 3.0 (-7)$^2$ \\
  SiS  &     2.89 (-7)    &   3.42 (-6)    &   1.59 (-7)    
&   1.36 (-9)    &   3.82 (-10)   & 4.4 (-7)$^1$ &  7.0 (-7)$^2$ \\
 {\bf  CH$_4$}  & {\bf    5.05 (-14) }   & {\bf   7.86 (-10) }   & {\bf  4.41 (-15) }    
& {\bf   8.59 (-20) }   & {\bf  1.24 (-20)}   & \\
  HS   &     6.49 (-6)    &   5.96 (-6)    &   1.06 (-6)     
&   1.63 (-8)    &   3.70 (-9)   & \\
  H$_2$S  &     1.03 (-6)    &   1.92 (-6)    &   2.15 (-8)    
&   3.95 (-12)    &   2.82 (-13)   & \\
  NS   &     4.36 (-11)    &   5.75 (-12)    &   3.68 (-11)    
&   1.91 (-12)    &   7.45 (-13)   & \\
  N$_2$   &     8.05 (-5)    &   2.76 (-6)    &   5.07 (-5)    
&   7.30 (-5)    &   6.90 (-5)   & \\
{\bf  HCN}  & {\bf    4.37 (-11) }   & {\bf   1.61 (-4)}    & {\bf   6.47 (-5)  }  
& {\bf   5.24 (-6)  }  & {\bf   2.12 (-6)}   & 9.8 (-7)$^1$ & 6.0 (-7)$^2$ \\
%%%  SiH$_4$ &     6.53 (-17)    &   6.40 (-18)    &   1.03 (-24)    
%%%&   1.45 (-27)    &   2.82 (-27)   & \\
  CN   &     1.73 (-13)    &   8.88 (-10)    &   6.06 (-10)    
&   4.75 (-10)    &   2.40 (-10)   & \\
%  S$_2$   &     7.67 (-9)    &   8.4 (-7)    &   7.95 (-7)
%&   9.33E-09    &   1.00E-09   & \\
{\bf   NH$_3$ } &  {\bf    1.20 (-10) }   &  {\bf  5.16 (-11)}    & {\bf   7.87 (-13)}    
&{\bf    3.33 (-16)}    &  {\bf  3.12 (-17) }  & $\sim$1.0 (-6)$^5$\\
  O    &     1.63 (-7)    &   9.41 (-12)    &   7.45 (-11)    
&   2.89 (-9)    &   1.11 (-9)   & \\
  OH   &     2.30 (-6)    &   1.10 (-8)    &   1.20 (-8)    
&   6.34 (-8)    &   5.53 (-8)   & \\
  H$_2$O  &     1.86 (-4)    &   3.68 (-4)    &   2.70 (-4)    
&   1.49 (-4)    &   1.02 (-4)   & \\
  CO   &     6.95 (-4)    &   5.50 (-4)    &   6.40 (-4)     
&   6.14 (-4)    &   5.38 (-4)   & \\
{\bf   CO$_2$}  & {\bf     4.09 (-8)}    &  {\bf  1.34 (-7) }   & {\bf   2.76 (-7) }   
& {\bf   3.39 (-5)  }  & {\bf   6.44 (-5)}   & \multicolumn{2}{c}{ISO detection$^4$}  \\
%  HCO  &     1.85 (-11)    &   2.20 (-11)    &   3.92 (-12)    
%&   4.05E-13    &   1.79 (-13)   & \\
  O$_2$   &     1.99 (-11)    &   1.73 (-15)    &   1.13 (-14)    
&   3.56 (-13)  &   1.25 (-13)   & \\
  SiO  &     4.28 (-5)    &   4.12 (-5)    &   4.43 (-5)    
&   4.06 (-5)    &   3.75 (-5)   & 1.7 (-5)$^1$ & 3.0 (-6)$^3$ \\
  NO   &     1.04 (-9)    &   1.70 (-13)    &   1.41 (-11)    
&   6.11 (-9)    &   1.46 (-8)   & \\
  SO   &     2.18 (-8)    &   4.87 (-9)    &   3.13 (-8)    
&   8.27 (-8)    &   7.79 (-8)   &  2.6 (-6)$^1$ & 1.8 (-6)$^3$ \\
  OCS  &     5.37 (-39)    &   1.60 (-9)    &   1.53 (-11)    
&   1.18 (-13)    &   2.52 (-14)   & \\
            \noalign{\smallskip}
            \hline
%%%         \end{array}
%%%      \] 
\end{tabular}
\end{flushleft}
\end{table*}

\section{Results and discussion}

The molecular abundances relative to the total gas number density are 
listed in Table 3 for various shock strengths and positions in the wind. 
As we do not know precisely the  position where the wind is chemically 
frozen due to the acceleration induced by grains, we have considered gas 
layers very close to the photosphere. Therefore, abundances at 2.2 
R$_\star$ may not be the exact values frozen in the outflow but are 
indicative of trends on destruction/formation of species in the inner 
wind. The dominant molecules  are, apart from molecular hydrogen, CO, H$_2$O, 
N$_2$ and SiO. This was known already from TE calculations applied to 
O-rich AGB stars and confirms the ``parent'' character of these 
molecules. However, and as for the case of carbon stars (see WC98), 
caution should be exerted in comparing observations with TE calculations. 
For some ``parent'' species, the non-equilibrium chemistry does not 
alter significantly the initial TE abundances.  However, other species 
abundant in the TE photosphere according to Table 3 (e.g. OH, O, SiS 
and HS) are quickly destroyed in the outflow by the non-equilibrium 
chemistry generated by shocks.   
 
This chemistry is also responsible for the formation of several 
carbon-bearing species close to the star, in particular CO$_2$, HCN, 
and CS. The theoretical values derived for HCN and CS are in excellent 
agreement with the abundances derived from millimeter lines in the 
outer envelope. Chemically, these species are quite stable and in 
carbon stars, they travel the entire envelope unaltered until they 
reach the photo-dissociation regions of the outer wind. The same 
should occur in O-rich winds as these molecules do not participate to 
the formation of dust grains (e.g., silicate, corundum) in the inner 
envelope. 
 
The chemical processes reponsible for the formation of HCN and CS 
are linked in that both CS and HCN are produced from reactions  
involving cyanogen, CN. While HCN is formed by the reaction
\begin{equation}
\hbox{CN} + \hbox{H$_2$} \rightarrow \hbox{HCN} + \hbox{H} 
\end{equation}
the formation of CS results from the reaction 
\begin{equation}
\hbox{CN} + \hbox{S} \rightarrow \hbox{CS} + \hbox{N}
\end{equation}
CN acts as an intermediate in the formation of the two molecules and 
is quickly destroyed by atomic hydrogen. Furthermore, the destruction 
of HCN by atomic hydrogen possesses a high activation barrier while 
Reaction (2) is highly dependent on temperature and then quite fast 
in the gas excursions.  

The chemical routes to the formation of CO$_2$ are 

\begin{equation}
\hbox{OH} + \hbox{CO} \rightarrow  \hbox{CO$_2$} + \hbox{H} 
\end{equation}
where hydroxyl OH is formed from the collisional destruction of water, 
H$_2$O, with atomic hydrogen, and
  
\begin{equation}
\hbox{CO} + \hbox{O} + \hbox{M} \rightarrow  \hbox{CO$_2$} + \hbox{M} 
\end{equation}

Reaction (4) represents the dominant formation pathway at small radii 
while Reaction (5) becomes important at larger radii, explaining the 
jump in the CO$_2$ abundance at 2 R$_\star$. The rate for Reaction (4) 
has been determined experimentally but that of Reaction (5) was calculated 
from thermodynamical data. In view of the uncertainty of this rate, we 
also consider Reaction (5) to proceed with a typical three-body reaction 
rate of k$= 2\times 10^{-32}$ cm$^6$ s$^{-1}$ to test the variation of 
CO$_2$ abundance with radius. The abundance was lower than that quoted 
in Table 3 but of the order of $\sim 10^{-6}$. Therefore, CO$_2$ is a 
direct result of shock chemistry involving the destruction of CO by OH radicals.  

Other species appear to be absent from the inner regions of the wind 
but are observed in the outer envelopes of Miras stars. This is the 
case for SO which is produced in the photo-dissociation 
regions by ion-molecule reactions according to WM97. However, WM97 
inject at large radii SiS when the molecule does not appear to be 
``parent'' and has a very low abundance in the inner wind (see Table 
3). This could explain the discrepancy found by WC97 between their 
theoretical value and the observed value for SiS which is much lower, 
implying that SiS is produced in the outer envelope of IK Tau. As for 
methane and ammonia, they are also absent from the inner wind of Mira 
stars, a result already derived for carbon stars (WC98). This points 
to the necessity of invoking formation processes different from pure 
gas-phase mechanisms and these species have to be produced in the 
intermediate regions of the wind from gas-grain interaction as first 
suggested by Nejad \& Millar (1988). Whether they enter the outer 
envelope of IK Tau with the large abundances used by WM97 needs to be 
confirmed by theoretical models or observations. The same conclusion 
is drawn for hydrogen sulphide, H$_2$S, which has a very low abundance at small 
stellar radii. Omont et al. (1993) have proposed a formation route for 
H$_2$S from gas-grain chemistry in the intermediate wind of Miras 
involving atomic sulphur and atomic hydrogen. We notice from Table 3 that both
S and H enter the intermediate part of the wind with quite
large abundances. This result for
sulphur could be tested observationally with the search of the [S I] fine
structure lines at 25 and 56 microns.

We conclude that the inner envelopes of O-rich Miras are regions of 
efficient formation of molecules and dust. As for carbon stars, the 
non-equilibrium chemistry is triggered by the propagation of periodic 
shocks induced by stellar pulsation. Some of the carbon molecules so 
far observed in these stars, in particular HCN, CS and CO$_2$ are 
formed in the inner wind and act as ``parent'' species throughout the 
envelope. This prediction awaits confirmation from new ISO data and 
sub-millimeter observations on O-rich Mira stars. 

\begin{acknowledgements} The authors wish to thank K. Justtanont for useful 
discussion and the referee, M. Lindqvist, for his helpful comments. 
\end{acknowledgements}

\end{document}